\documentclass[showpacs,amsmath,pre]{revtex4}
\usepackage{graphicx}
\usepackage{amsmath}

\usepackage{epsfig}
\usepackage{graphics}
\usepackage{latexsym}
\usepackage{amsfonts}
\usepackage{amssymb}
   
\usepackage{plain}

\begin{document}

\title{Gas adsorption/desorption  in silica aerogels: a theoretical study of scattering properties}

\author{F. Detcheverry, E. Kierlik, M. L. Rosinberg, and G. Tarjus}
\affiliation{Laboratoire de Physique Th\'eorique de la Mati\`ere Condens\'ee, Universit\'e Pierre et
Marie Curie, 4 place Jussieu, 75252 Paris Cedex 05, France}
  

\begin{abstract}
We  present a numerical study of the structural correlations  associated to gas
adsorption/desorption  in silica aerogels in order to provide a theoretical
interpretation of 
scattering experiments. Following our earlier work, we use a coarse-grained lattice-gas description
and determine the nonequilibrium behavior of the adsorbed gas within a local mean-field analysis.
 We focus on the differences between the adsorption and desorption mechanisms and
 their signature in the fluid-fluid and gel-fluid structure factors as
a function of temperature. At low temperature, but still in the regime where the isotherms are continuous, we find that  the adsorbed
fluid density, during both filling and draining, is correlated over distances that may be much larger than the  
gel correlation length.
In particular, extended fractal
correlations may occur during desorption, indicating the existence of  a ramified cluster of vapor filled
cavities. This also induces an important increase of the scattering intensity at small wave
vectors. The similarity
and differences with the scattering of fluids in other porous solids such as Vycor are discussed.

\end{abstract}
\maketitle

\section{Introduction}
  
In a series of recent papers\cite{DKRT2003,DKRT2004,DKRT2005}, we have presented a 
comprehensive theoretical study of gas adsorption/desorption in 
silica aerogels, revealing the  microscopic mechanisms that underly the
changes in the morphology of the hysteresis loops with temperature and gel porosity.
In particular, we have shown that the traditional  capillary condensation scenario
based on the independent-pore model\cite{GS1982} does not apply to aerogels, as a
consequence of the ``open" nature and interconnectedness of their microstructure. We have
found, on the other hand, that nonequilibrium phase transitions  (that differ on adsorption
 and desorption) are at the origin of the very steep isotherms
 observed with $^4$He in high porosity gels at low temperature\cite{C1996,TYC1999}.  In this  work, we complete our 
 study by investigating the  correlations within the adsorbed fluid and computing the 
 fluid-fluid and solid-fluid structure factors that can be  measured
 (at least indirectly) in scattering experiments.  Scattering methods
 (using x-rays, neutrons and visible light) are now frequently combined with thermodynamic
 measurements for extracting information on the structure and the dynamics of the
 adsorbed molecules and understanding the influence of solid microstructure
 on fluid properties\cite{H2004}. In the case of $^4$He in aerogel, both small-angle x-ray scattering (SAXS)\cite{LMPMMC2000} 
 and light scattering  measurements\cite{LGPW2004} have been recently performed along the sorption isotherms.
 However, the interpretation of the scattered
 intensity is always complicated by the fact that  it contains several contributions that
 cannot be resolved without assuming some mechanism for the sorption
 process. For instance, in the case of a low porosity solid like Vycor, the evolution of the scattered intensity along the capillary rise 
 is usually interpreted in the framework of an independent-pore model, with the gas condensing
 in pores of increasing size that are (almost) randomly distributed throughout the material\cite{LRHHI1994,PL1995,KKSMSK2000}. This  
 explains that  long-range correlations are not observed during adsorption. On the other hand, we have shown that large-scale
 collective events occur in aerogels, and this may have a significant influence on the scattering properties. Indeed, we shall
 see in the following that scattering curves at low temperature may not reflect  the underlying
 microstructure of the gel.  More generally, our main objective is to understand 
  how the different
 mechanisms  for adsorption and desorption reflect in the scattering properties as the temperature is changed. (Note that there has been
  a recent theoretical study of this problem that is closely related to the present one\cite{SG2004}; there are, however, 
 significant differences that will be commented in due place.)
 Particular attention will be paid to the `percolation invasion' regime that is predicted
 to occur during the draining of gels of medium porosity (e.g. $87\%$) and that
 manifests itself by the presence of fractal correlations. Such correlations have been 
 observed in Vycor\cite{LRHHI1994,PL1995,KKSMSK2000} and xerogel\cite{H2002}, but no experiment has been carried out so far to
 detect a similar phenomenon in aerogels. We therefore hope that the present work will
 be an incentive for such a study.
On the other hand, the influence of gel porosity on scattering properties will only be
 evoked very briefly. In particular, for  reasons that will be 
explained below, the correlations along the steep (and possibly discontinuous due to nonequilibrium phase transitions)
isotherms observed in high porosity gels at low temperature are not  investigated.

The paper is organized as follows. In section 2, the model and the theory
are briefly reviewed and the cart.texomputation of the correlation functions and the corresponding
structure factors is detailed. The numerical results are presented in
section 3. The relevance of our results to existing and future scattering experiments is
discussed in section 4.

\section{Model and correlation functions}

\subsection{Lattice-gas model}

As discussed in previous papers\cite{DKRT2003,DKRT2004,DKRT2005},
our approach is based on a coarse-grained lattice-gas description
which incorporates the essential physical ingredients of
gel-fluid systems. The model Hamiltonian is given by
\begin{align}
{\cal H} = -&w_{ff}\sum_{<ij>} \tau_{i}\tau_{j} \eta_i \eta_j
-w_{gf}\sum_{<ij>
}[\tau_{i}\eta_i (1-\eta_j)+\tau_{j}\eta_j (1-\eta_i)]-\mu \sum_i \tau_i\eta_i
\end{align}
where  $\tau_i=0,1$ is the fluid occupation variable ($i=1...N$)
and $\eta_i=1,0$ is the quenched random variable that describes the solid microstructure (when $\eta_i=0$, site $i$
is  occupied by the gel; $\phi=(1/N)\sum_i\eta_i$ is thus the gel porosity). Specifically, we adress the case of base-catalyzed silica
aerogels (typically used in helium experiments) whose structure is well accounted for by a
diffusion limited cluster-cluster aggregation algorithm (DLCA)\cite{M1983}. In the Hamiltonian, 
$w_{ff}$ and $w_{gf}$ denote respectively the fluid-fluid and
gel-fluid attractive interactions,  $\mu $ is the fluid chemical
potential (fixed by an external reservoir), and the double summations  run over all distinct pairs of
nearest-neighbor (n.n.) sites.

Fluid configurations along
the sorption isotherms are computed using local mean-field theory (i.e. mean-field density
functional theory), neglecting
thermal fluctuations and activated processes (the interested reader is referred to
Refs.\cite{DKRT2003,DKRT2004,DKRT2005} for a detailed presentation of the
theory). As $\mu$ varies, the system visits a sequence of
metastable states which are local minima of the following grand-potential functional
\begin{align}
\Omega(\{\rho_i\})&=k_BT \sum_i[\rho_i\ln \rho_i+(\eta_i-\rho_i)\ln(\eta_i-\rho_i)]
-w_{ff} \sum_{<ij>}\rho_i\rho_j 
-w_{gf}\sum_{<ij>}[\rho_i(1-\eta_j)+\rho_j(1-\eta_i)] -\mu\sum_i\rho_i\
\end{align}
where $\rho_i(\{\eta_i\})=<\tau_i\eta_i>$ is the thermally averaged fluid density at site $i$.
Earlier work has shown that this approach reproduces qualitatively
the main features of adsorption phenomena in disordered porous solids\cite{KMRST2001,SM2001}.

All calculations presented below were performed on a body-centered cubic lattice of
linear size $L=100$ ($N=2L^3$) with periodic boundary conditions in all directions (the 
lattice spacing $a$ is taken as the unit length).
$L$ is large enough to faithfully describe gels with porosity $\phi\leq 95\%$, but
an average over a significant number of gel realizations is required to obtain a
good description of the correlation functions. In the following,
we use $500$ realizations. $w_{ff}$ is taken as the energy unit and temperatures are expressed in the reduced unit $ T^*=T/T_c$, where $T_c$ is the critical
temperature of the pure fluid ($T_c=5.195K$ for helium and $kT_c/w_{ff}=2$ in the theory).
The interaction ratio $y=w_{gf}/w_{ff}$ is equal
 to $2$ so as to reproduce approximately the height of the hysteresis loop
 measured with $^4$He in a $87\%$ porosity aerogel at $T=2.42K$\cite{DKRT2003,TYC1999}.
\subsection{Correlation functions and structure factors}

In a scattering experiment performed in conjunction with gas adsorption/desorption, one
typically measures a (spherically averaged)
scattered intensity $I(q)$ which is proportional to a combination of the
three partial structure factors $S_{gg}(q), S_{ff}(q)$ and $S_{gf}(q)$, where $g$ and $f$
denote the gel and the fluid, respectively.
Whereas a macroscopic sample is usually considered as
isotropic and statistically homogeneous, our calculations are
performed in finite samples and on a lattice, and some work is
needed to obtain structure factors that can be possibly compared to experimental data.

Let us first consider the gel structure factor. The calculation of
$S_{gg}(q)$ proceeds in several steps. As in the case of an off-lattice DLCA
simulation\cite{H1994}, we first compute the two-point
correlation function $g_{gg}({\bf r})=h_{gg}({\bf r})+1$ by performing a double 
average over the lattice sites and the gel realizations,
\begin{align}
\rho_g^2 g_{gg}({\bf r}) = \frac{1}{N}\sum_{i,j} \overline{(1-\eta_i)(1-\eta_j)}
\delta_{{\bf r},{\bf r}_{ij}}-\rho_g \delta_{{\bf r},{\bf 0}}
\end{align}
taking care of the periodic boundary conditions. Here, ${\bf r}_{ij}={\bf r}_{i}-{\bf r}_{j}$, $\rho_g=1-\phi$ is the lattice fraction occupied by the gel, and the second term in the right-hand side
takes into account the fact that there are only one particle per site, which
yields the (point) hard-core condition, $g_{gg}({\bf r=0})=0$.  
In this expression  and in the following, the overbar denotes an average over different gel
realizations produced by the DLCA algorithm. The computation of $g_{gg}({\bf r})$
is efficiently performed by introducing the Fourier transform
$\rho_g({\bf q})=\sum_i (1-\eta_i)\exp(-2i\pi{\bf q.r_i}/N)$
where ${\bf q}$ is a vector of the reciprocal lattice, and by
using the fast Fourier transform (FFT) method\cite{MLW1996,SG2004}. This reduces the
computational work to $O(N \ln N)$ instead of $O(N^2)$ when the direct real-space
route is used. (The same method is applied to the other correlation functions.)

In a second step, we ``sphericalize" the correlation function by collecting the values having same
modulus of the argument ${\bf r}$,
\begin{align}
g_{gg}(r)=\frac{ \sum_{{\bf r}'} g_{gg}({\bf r}') \delta_{r,r'}}
{ \sum_{{\bf r}'} \delta_{ r,r'}} \  .
\end{align}
Finally, instead of storing the values of $g_{gg}(r)$ for all possible distances $r_{ij}$ on the lattice between
 $d=a\sqrt{3}/2$, the
 nearest-neighbor distance, and $L/2$, we  bin the data with a spacing $\Delta r=0.05 $ and
 interpolate  linearly between two successive points (the restriction to
 $r<L/2$ avoids boundary artefacts). Moreover, we impose the
 ``extended" hard-core condition $g_{gg}(r)=0$ for $r<d$, in line with our interpretation of
 a gel site as representing an impenetrable silica particle\cite{DKRT2003}.
 (In Ref. \cite{SG2004}, in contrast, the model is thought of as a discretization of
 space into cells of the size of a fluid molecule and the gel particle ``radius"  is varied
 from $2$ to $10$ lattice spacings.)
  Of course, the interpolation procedure does not completely erase
 the dependence on the underlying lattice structure, especially at short distances.
Following Ref.\cite{H1994}, the structure factor is then computed using
\begin{align}
S_{gg}(q)= 1+ 4\pi\rho_g \int_0^{L/2} r^2 ( h_{gg}(r)-h_{gg}) \frac{\sin(qr)}{qr} dr
\end{align}
where $h_{gg}$ is a very small parameter adjusted such that
$S_{gg}(q)\rightarrow 0$ as $q\rightarrow 0$. Indeed, since the
DLCA aerogels are built in the ``canonical" ensemble with
a fixed number of gel particles $N_g=\rho_g N$, the following sum-rule holds:
\begin{align}
\rho_g\sum_{{\bf r}}g_{gg}({\bf r}) = N_g-1
\end{align}
which readily  yields $S_{gg}(0)=1+\rho_g\sum_{{\bf r}} h_{gg}({\bf r})=0$ in Fourier space.
This trick allows one to obtain  a reasonable continuation of $S_{gg}(q)$
below $q_{min}=2\pi/L$\cite{H1994}.

Similarly, the fluid-fluid two-point correlation function 
$g_{ff}({\bf r})=1+h_{ff}({\bf r})$ is defined as
\begin{align}
\rho_f^2 g_{ff}({\bf r})=\frac{1}{N}\sum_{i,j}\overline{\langle\tau_i\eta_i\tau_j\eta_j\rangle}
\delta_{{\bf r},{\bf r}_{ij}}-\rho_f \delta_{{\bf r},{\bf 0}}
\end{align}
where  $\rho_f=(1/N)\sum_i\overline{\langle\tau_i\eta_i\rangle}=(1/N)\sum_i\overline{\rho_i}$ is the 
average fluid density and the sum is performed again over all lattice
sites to improve the statistics (for notational
simplicity, we have dropped the overbar on $\rho_f$).
Because of the double average over thermal fluctuations and over disorder, there
are two distinct contributions to $h_{ff}({\bf r})$, which are usually called ``connected"
and ``blocking" or ``disconnected" \cite{GS1992,RTS1994}, and which, in the present case, are given by 
the expressions,
\begin{align}
&\rho_f^2 h_{ff,c}({\bf r})+\rho_f \delta_{{\bf r},{\bf 0}}=
\frac{1}{N}\sum_{i,j}\overline{[\langle\tau_i\eta_i\tau_j\eta_j\rangle
-\rho_i\rho_j]} \ \delta_{{\bf r},{\bf r}_{ij}}\\
&\rho_f^2 h_{ff,d}({\bf r})=\frac{1}{N}\sum_{i,j}[\overline{\rho_i\rho_j}-\rho_f^2]\
\delta_{{\bf r},{\bf r}_{ij}}  \ .
\end{align}
In the pure fluid, $h_{ff,c}({\bf r})$ is just the standard connected pair 
correlation function whereas $h_{ff,d}({\bf r})$ has no equivalent. It turns out, however, that only
$h_{ff,d}({\bf r})$ can be
computed along the sorption isotherms. Indeed, the quantity $\langle\tau_i\eta_i\tau_j\eta_j\rangle$ cannot be obtained in the framework of  mean-field theory, and the only available route to  $h_{ff,c}({\bf r})$ is via the ``fluctuation" relation\cite{RTS1994}
\begin{align}
\rho_f^2 h_{ff,c}({\bf r}_{ij})+\rho_f\delta_{{\bf r}_{ij},{\bf 0}}=
\frac{\partial^2 \beta \Omega}
{\partial (\beta \mu_i )\partial (\beta \mu_j)}=\frac{\partial\rho_i}{ \partial (\beta \mu_j)}
\end{align}
where $\mu_i$ is a site-dependent chemical potential\cite{note2}.
However, this relation only holds  at equilibrium (like the Gibbs adsorption equation
$\rho_f=-(1/N)\partial \Omega/ \partial \mu$  discussed in Ref.\cite{DKRT2003}) and therefore
it cannot be applied along the hysteresis loop where the system jumps from one
metastable state to another. (In a finite sample, the grand potential changes discontinuously
along the adsorption and desorption desorption branches\cite{DKRT2003}.) We are thus forced to approximate
$h_{ff}({\bf r})$  by its disconnected part, $h_{ff,d}({\bf r})$\cite{note1}.
However,  this may not be a bad approximation at low temperature
 because the
local fluid densities $\rho_i$ are then very close to zero or one\cite{DKRT2003}, which likely implies that
$h_{ff,c}({\bf r})$ is a much smaller quantity than $h_{ff,d}({\bf r})$\cite{note4}.

We then apply to $h_{ff}({\bf r})$ the same procedure as for $g_{gg}({\bf r})$, taking
radial averages and then performing a binning of the data and a linear interpolation.
There are no ``extended" hard-core in this case. Indeed, since  the scale of
the coarse-graining is fixed by the size of a gel particle (typically, a few nanometers), a
lattice cell may contain several hundreds of fluid molecules which may be thus 
considered as point particles.
$h_{ff}(r)$ is then  also interpolated between $r=0$ and $r=d$.

In a grand-canonical calculation, the number of fluid particles  fluctuates from sample to sample, which implies the following sum-rule for the disconnected pair correlation function $h_{ff,d}({\bf r})$\cite{RTS1994} (and thus for $h_{ff}({\bf r})$ in our approximation)
\begin{equation}
\sum_{\bf r} h_{ff}({\bf r}) \simeq\sum_{\bf r}h_b({\bf r}) =N \frac{\overline{{\rho_f^2(\{\eta_i\})}}-\rho_f^2}{\rho_f^2}
\end{equation}
where $\rho_f(\{\eta_i\})=(1/N)\sum_i \rho_i$ is the average fluid density for a 
given gel realization. This sum-rule can be used to also extrapolate $S_{ff}(q)$ below $q<2\pi/L$, using
\begin{align}
S_{ff}(q)= 1+ 4\pi\rho_f  \int_0^{L/2} r^2 ( h_{ff}(r)-h_{ff}) \frac{\sin(qr)}{qr} dr
\end{align}
where $h_{ff}$ is adjusted so that 
$S_{ff}(0)=1+ N [\overline{\rho_f^2(\{\eta_i\})}-\rho_f]/\rho_f$.

Finally, the gel-fluid two-point correlation function $g_{gf}({\bf r})=1+h_{gf}({\bf r})$ is computed from
\begin{align}
\rho_g \, \rho_f \, g_{gf}({\bf r})=\frac{1}{N}\sum_{i,j}\overline{(1-\eta_i)\langle\tau_j\eta_j\rangle}
\delta_{{\bf r},{\bf r}_{ij}}
\end{align}
and then sphericalized, binned, and linearly interpolated (taking $g_{gf}(r)=0$
for $0<r<d/2$ since no fluid molecule can be found at a distance less than $d/2$
from the centre of a gel particle).
The cross structure factor $S_{gf}(q)$ is then obtained from
\begin{align}
S_{gf}(q)=  4\pi \sqrt{\rho_g\rho_f} \int_0^{L/2} r^2 ( h_{gf}(r)-h_{gf}) \frac{\sin(qr)}{qr} dr
\end{align}
where $h_{gf}$ is adjusted so as to satisfy the sum-rule $S_{gf}(q\rightarrow0)=0$ 
which again results from the absence of fluctuations in the number of gel particles.

\section{Numerical results}

\subsection{Aerogel structure}

We first concentrate on the case of the empty aerogel. We have already presented in  Ref.\cite{DKRT2003}  the pair correlation function $g_{gg}(r)$  for
several porosities between $87\%$ and $99\%$.   These curves exhibit a shallow minimum that strongly depends on $\phi$
and whose position gives an estimate of the gel correlation lenght $\xi_g$, as 
suggested in Ref.\cite{H1994}. A DLCA  gel can
 indeed be sketched as a disordered packing of ramified blobs with average size 
 $\xi_g$. For instance, $\xi_g$ varies approximately from $4$ to $10$ lattice spacings
 as the porosity increases from
 $87\%$ to $95\% $ (this is much smaller than the box size $L$, which ensures that 
 there are no finite-size artifacts in the calculation of the gel structure\cite{note5}).
Only  the highest-porosity samples exhibit a significant power-law regime 
$g_{gg}(r)\sim r^{-(3-d_f)}$
that reveals the fractal character of the intrablob correlations.
\begin{figure}[hbt]
\includegraphics*[width=9cm]{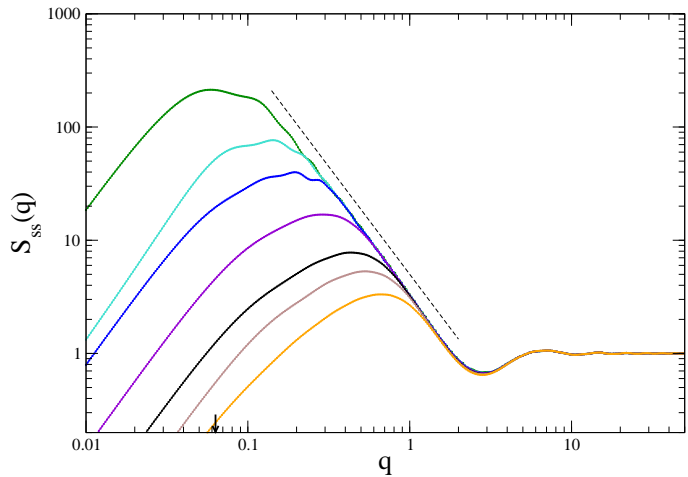}
\caption{Gel structure factors $S_{gg}(q)$ obtained with the DLCA algorithm for different
porosities. From left to right: $\phi=0.99, 0.98, 0.97, 0.95, 0.92, 0.90, 0.87$. The dashed line has a slope $-1.9$. The arrow indicates the wavevector $q=2\pi/L\approx 0.063$.  (Color on line).}
\end{figure}
Although we shall essentially focus in the following on the case of the $87\%$ porosity gel,
 for the sake of completeness we show in Fig. 1
the evolution of the simulated gel structure factor with porosity.

The curves closely resemble those obtained with  the continuum model\cite{H1994}.
In particular, they exhibit the same damped oscillations  at large $q$ that result from
the ``extended" hard-core condition $g_{gg}(r)=0$ for $r<d$ (the oscillations, however, are more significant in the continuum model). The range of the
linear fractal regime increases with porosity (it is almost non-existent in the $87\%$ gel) and corresponds asymptotically to
a fractal dimension $d_f\approx 1.9$ (the value $d_f\approx 1.87$ was obtained
in Ref.\cite{DKRT2003} from the $g_{gg}(r)$ plot for the $99\%$ aerogel). A characteristic feature of the curves is the existence of a
maximum at smaller wavevectors whose location  $q_{m}$ decreases with porosity and correlates well with
$1/\xi_g$ ($q_{m} \sim 2.6/\xi_g$). This  maximum is thus the Fourier-space signature  of
the shallow minimum observed in $g_{gg}(r)$. (Note that varying the size $L$ has only a weak influence on the small $q$ 
portion of the curves for the  $87\%$ and $95\%$ gels, which validates the continuation procedure used in Eq. 5.)

To compute the scattering intensity $I(q)$ and compare to the results of small-angle 
x-rays or neutron experiments, it is necessary to introduce a form factor $F(q)$ 
for the gel particles. One can use for instance the form factor of spheres with radius $R=d/2$, 
$F(q)=3 [\sin(qR)-qR\cos(qR)]/(qR)^3$. The curve $I(q)=S_{gg}(q)F(q)^2$ then
differs from $S_{gg}(q)$  in the large-$q$  regime  ($q/2\pi>d^{-1}$) where it follows the 
Porod law $I(q)\sim q^{-4}$\cite{P1982}.  On the other hand, the intermediate 
``fractal" regime ($\xi_g^{-1}<q/2\pi<d^{-1}$) where $S_{gg}(q)\sim q^{-d_f}$, and
the location $q_{m}$  of the maximum are essentially unchanged.  By comparing the 
value of $q_m$ in Fig. 1 with the actual value in the experimental curves
(displayed for instance  in Ref.\cite{H1994}), we can thus fix approximately the
scale of our coarse-graining. For $\phi=95\%$, $q_m\approx 0.01 $\AA{}$^{-1}$ which 
yields $a\approx 3$ nm, a reasonable value for base-catalysed silica aerogels which is in agreement 
with the estimation of Ref.\cite{DKRT2003} obtained from the gel correlation 
length $\xi_g$.
It is worth noting that the DLCA simulated structure factors present a more pronounced maximum at $q_m$ than  the experimental  curves $I(q)$, as already noticed in the literature\cite{H1994,OJ1995}. There are obviously large scale inhomogeneities in actual aerogels that are not reproduced in the simulations.  Moreover, as emphasized in Refs.\cite{H1994,OJ1995}, there are some significant  details that are neglected in the DLCA model, such as the rotational diffusion of the aggregates, their polydispersity and irregular form, and all kinds of possible restructurating effects.

\subsection{Fluid structure during adsorption}
\begin{figure}[hbt]
\includegraphics*[width=8cm]{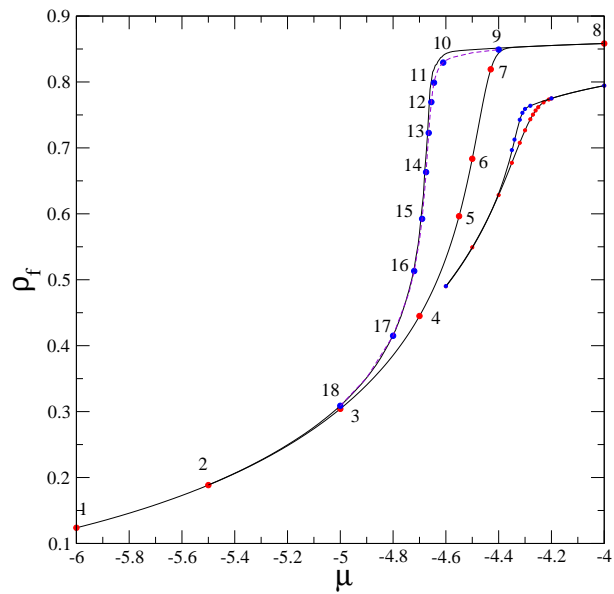}
\caption{Average hysteresis loops in a $87\%$ porosity aerogel at $T^*=0.5$ and $0.8$
(from left to right).
 The points along the adsorption and
desorption isotherms at $T^*=0.5$
 indicate the values of the chemical potential for which the correlation 
 functions are computed. The desorption isotherm has been computed either in presence of an external reservoir (solid line) or by using the procedure described  in section IIIC (dashed line)  (Color on line).}
\end{figure}
As shown
in Refs.\cite{DKRT2003,DKRT2004,DKRT2005},  the elementary condensation events 
(avalanches) that occur in 87\% porosity aerogel as the chemical potential is slowly varied
are always of microscopic size, whatever the temperature. This implies that the 
adsorption isotherms are smooth in the thermodynamic limit or when averaging over 
a large number of finite samples, as illustrated in
Fig. 2.
We have computed the correlation functions and the corresponding structure factors
for a number of points along the $T^*=0.5$ and $T^*=0.8$ isotherms, as indicated in
the figure. We first consider the lowest temperature.

Figs. 3 and 4  show the evolution of the correlation functions $h_{ff}(r)$  and  
$h_{gf}(r)$ with chemical potential.
One can see that the curves change significantly  as $\mu$ increases
(note that the vertical scale in Fig. 3(a) is expanded so as to emphasize the presence of
a shallow minimum in the curves; accordingly, the values of $h_{ff}(r)$ near zero are not visible).
For very low values of the chemical potential (e.g. $\mu=-6$), $h_{ff}(r)$ looks 
very much like
$h_{gg}(r)$, apart from a shift towards larger values of $r$ by a distance of about 
$2$ lattice spacings. Indeed, as shown in our
earlier work\cite{DKRT2003,DKRT2004}, in the early stage of  the adsorption process, 
the adsorbed fluid forms a liquid film that coats the aerogel strands and whose 
thickness is approximately one lattice spacing at low temperature. In consequence, the distribution
of the fluid particles follows the spatial arrangement of the aerogel, a feature
already observed in a continuum version of the model\cite{KKRT2001}.
The existence of the liquid film also reflects in the rapid decrease of $h_{gf}(r)$ 
with $r$, which indicates that the fluid is only present in the vicinity  of the gel 
particles (the fact that $h_{gf}(d)<h_{gf}(a)$ may be ascribed to the connectivity 
of the gel: around a gel particle, there are always other gel particles - $2.5$ in average  in the first shell - and the probability to find a fluid particle is thus suppressed).
\begin{figure}[hbt]
\includegraphics*[width=12cm]{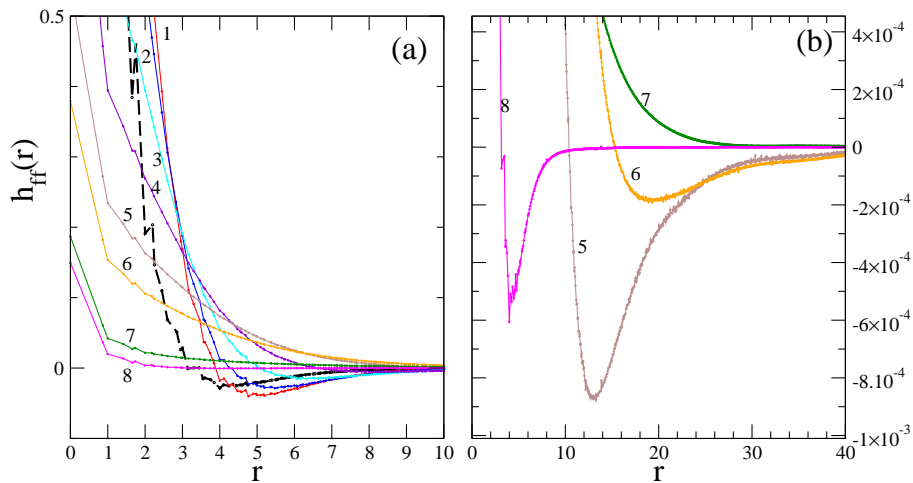}
\caption{ Fluid-fluid correlation function $h_{ff}(r)$ along the adsorption isotherm 
in a $87\%$ porosity aerogel at $T^*=0.5$. (a): From top to bottom, the curves correspond to points $1$ to $8$ in Fig. 2; the dashed line is the gel correlation function $h_{gg}(r)$. (b): Magnification of (a) showing the evolution of the minimum as $\mu$ varies from  $-4.55$ to $-4$ (points $5$ to $8$ in Fig. 2) (Color on line).}
\end{figure}
\begin{figure}[hbt]
\includegraphics*[width=12cm]{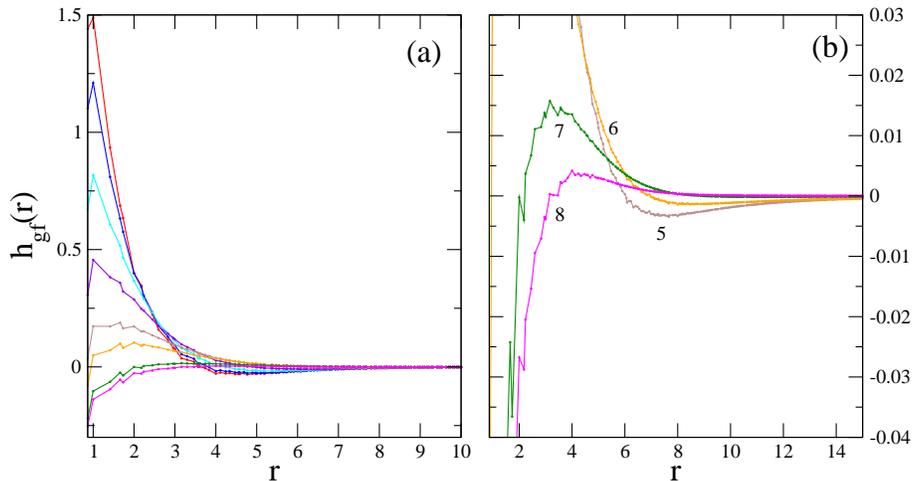}
\caption{ Same as Fig. 3  for the cross correlation function $h_{gf}(r)$ (Color on line).}
\end{figure}
As $\mu$ increases, the magnitude of the fluid-fluid correlations decreases at small $r$
(the contact value decreases from $4.45$ to $0.15$) and the depth of the minimum
in $h_{ff}(r)$ decreases as it shifts to larger values of $r$
(its location varies from $5$ to $24$ as  $\mu$ increases
from $-6$ to $-4.47$). The minimum disappears as the last voids
in the gel fill with liquid (note the difference in the vertical
scales of Figs. 3(a) and 3(b)), and finally, as one reaches saturation 
(at $\mu_{sat}=-4$), the shape of the gel-gel correlation function $h_{gg}(r)$ 
is approximately recovered, in accordance with
Babinet principle\cite{P1982,note6}. A similar behavior is observed
in the cross correlation function $h_{gf}(r)$ in Fig. 4, but the minimum occurs at a
smaller distance (its location is approximately $r\approx 12$ when it disappears).

Like for $g_{gg}(r)$, we may associate to the location of the mimimum in $h_{ff}(r)$ a length $\xi_{f}$ that
characterizes the correlations within the adsorbed fluid. The fact that $\xi_{f}$
becomes significantly larger than $\xi_g$ as the adsorption proceeds
shows that the fluid develops its own complex structure that does not reflect anymore 
the underlying gel structure. This is likely in relation with 
the fact that some of the condensation events studied in our
previous works\cite{DKRT2004,DKRT2005} extend much beyond the largest voids in 
the aerogel\cite{note9}.  It is worth noting that
these large avalanches (with a radius of gyration
$R_g\approx 12$\cite{DKRT2004,DKRT2005})
occur approximately in the same range of chemical
potential ($-4.5\leq\mu \leq-4.4$) where $\xi_f$ reaches its maximum (this also corresponds to the steepest portion of the
isotherm). 

\begin{figure}[hbt]
\includegraphics*[width=10cm]{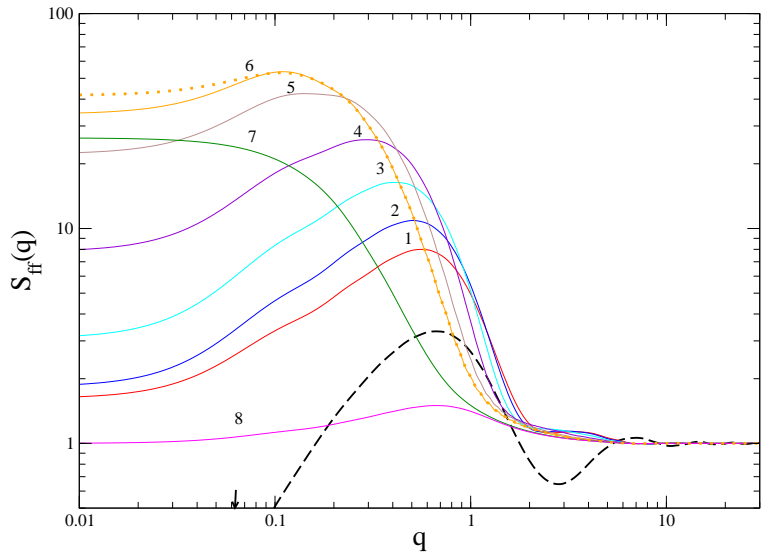}
\caption{ Fluid-fluid structure factor  $S_{ff}(q)$ along the adsorption isotherm
in a $87\%$ porosity aerogel at $T^*=0.5$. The numbers refer to points $1$ to $8$
in Fig. 2. The dashed line is the gel structure factor $S_{gg}(q)$ and the dotted line
illustrates the influence of the continuation procedure for $q<2\pi/L$ 
(see \cite{note7}) (Color on line).}
\end{figure}
\begin{figure}[hbt]
\includegraphics*[width=10cm]{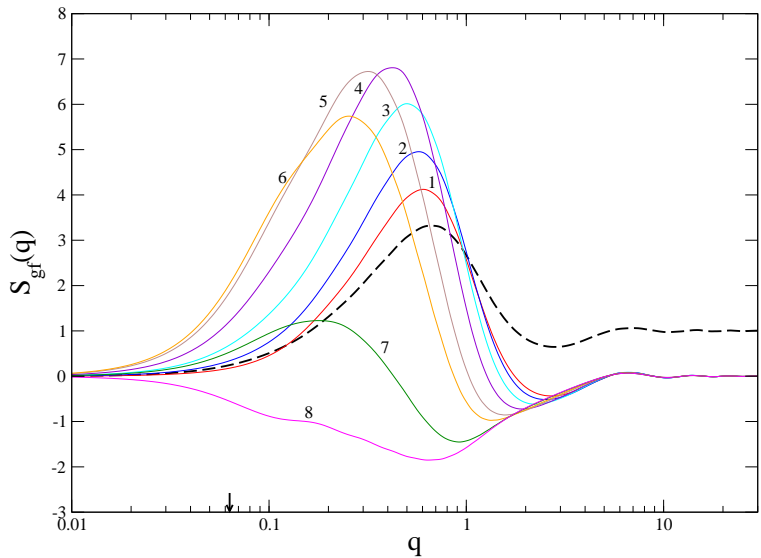}
\caption{Same as Fig. 5  for the cross structure factor $S_{gf}(q)$. Note that the vertical
scale is not logarithmic because $S_{gf}(q)$ has negative values (Color on line).}
\end{figure}

The corresponding structure factors $S_{ff}(q)$ and $S_{gf}(q)$ are shown in
Figs. 5 and 6, respectively\cite{note7}. The main feature in $S_{ff}(q)$ is the presence of a broad peak that grows and moves towards smaller wavevector as the fluid condenses in the
porous space. This peak is clearly associated to the minimum in $h_{ff}(r)$  (its location is approximately proportional
to $\xi_f^{-1}$) and it thus tells us the same story: the growing of a characteristic
length scale in the fluid along the capillary rise. The peak disappears in the last 
stage
of the adsorption process and is then replaced by a plateau (see curve 7 in Fig. 5). Finally, at
$\mu=\mu_{sat}$, one recovers a structure factor that can be deduced from $S_{gg}(q)-1$
by a linear transformation, in accordance with
Babinet principle (there are no oscillations in $S_{ff}(q)$, however, because
the fluid-fluid hard-core diameter is zero).

The evolution of the gel-fluid cross structure factor $S_{gf}(q)$ is somewhat different.
The peak is more prominent, as a consequence of the `no-fluctuation' condition
$S_{gf}(q=0)=0$ (this
feature may not be so pronounced in actual systems because of large-scale 
fluctuations), and it is located at a larger wavevector than in $S_{ff}(q)$ (in line with
the corresponding locations of the minima in $h_{ff}(r)$ and $h_{gf}(r)$). The most substantial
difference with $S_{ff}(q)$ is that the amplitude of the peak starts to decrease much before
the end of  the adsorption
process. The negative correlation observed at saturation is again due to
complementarity with the gel-gel structure\cite{SG2004}.

We have repeated these calculations at $T^*=0.8$ in order to investigate the
influence of temperature. In the $87\%$ gel, $T^*=0.8$ is just below $T_h$,
the temperature at which the hysteresis loop disappears (see Fig. 2).
$h_{ff}(r)$ and $h_{gf}(r)$ still exhibit a minimum that moves towards larger $r$
upon adsorption. However, the characteristic length $\xi_f$, associated to the 
minimum of $h_{ff}(r)$, does not exceed $14$ lattice spacings, indicating that
the size of the inhomogeneities in the fluid decreases with increasing $T$. A similar observation
was made in Ref.\cite{DKRT2004,DKRT2005} concerning the size of the avalanches
which become more compact at higher temperature and often
correspond to a condensation event occuring in a single cavity of the aerogel.
The shape of the corresponding structure factors does not change significantly with respect to the $T^*=0.5$ case,
but the amplitude is significanly reduced: the maximal amplitudes of the peaks in $S_{ff}(q)$ and $S_{gf}(q)$
are divided approximately by $5$ and $2$, respectively.

As shown in Refs.\cite{DKRT2003,DKRT2004,DKRT2005}, temperature has a much more
dramatic influence on the adsorption process in gels of higher porosity. In particular, 
at low enough temperature ($T<T_c(\phi)$ with $T_c^*(\phi)\approx 0.5$ in the $95\%$ gel\cite{DKRT2005}), a macroscopic 
avalanche occurs at a certain value of the chemical potential, with the whole sample
filling abruptly, which  results in a discontinuous isotherm
in the thermodynamic limit. In a finite system, the signature of a macroscopic avalanche is 
a large jump in the fluid density whose location in $\mu$ fluctuates from sample to sample,
which results in a steep but smooth isotherm after an average over the gel realizations (one then has 
to perform a finite-size scaling study to conclude on the proper behavior in the thermodynamic limit\cite{DKRT2003,DKRT2005}).
Within a grand canonical calculation, there is unfortunately no way to study the evolution of the structural
properties of the fluid during a macroscopic avalanche as this would require to consider intermediate fluid densities that
are inaccessible\cite{note8} (the situation would be different if the fluid density
was controlled instead of the chemical potential, as is done frequently in  experiments\cite{TYC1999,WC1990}). All we can do is to study
the $95\%$ gel at a higher temperature where the adsorption is still gradual, for instance at $T^*=0.8$.  In this case, no 
qualitative change is found  with respect to the case of the $ 87\%$  gel at $T^*=0.5$.  Indeed, as emphasized in Ref.\cite{DKRT2005},
adsorption proceeds similarly in a high-porosity gel at high temperature and in a lower-porosity gel at low temperature.  The 
correlation length $\xi_f$ is somewhat  larger in the $95\%$ gel (beyond $30$ lattice spacings)  so that finite-size effects come into play (in particular,
it becomes problematic  to extraplotate $S_{ff}(q)$ to $q=0$ so as to simulate the infinite-size limit). To go to lower temperatures, it 
would be thus necessary to use a much larger simulation box, which would increase considerably the computational work. Note that   one expects  $h_{ff}(r)$ to decay algebraically at the critical temperature $T_c(\phi)$. Indeed, according to the analogy with  the $T=0$ nonequilibrium random-field Ising model (RFIM), there should be only one important  length scale in the system close to criticality, length scale which is proportional to the  average linear extent of the largest avalanches\cite{DS1996}. At criticality, this correlation length diverges.

\subsection{Fluid structure during desorption}

As discussed in detail in Refs.\cite{DKRT2003,DKRT2004}, different mechanisms may be responsible for gas desorption in aerogels, depending
on porosity and temperature. In contrast with adsorption, the outer surface of the material where the adsorbed
fluid is in contact with the external vapor may play an essential role. 
For instance, in the $87\%$ aerogel at $T^*=0.5$, the theory predicts
a phenomenon akin to percolation invasion : as $\mu$ is decreased from saturation, 
some gas ``fingers"  enter the sample and grow until they percolate at a certain 
value of $\mu$, forming a fractal, isotropic cluster. The desorption then proceeds gradually via the growth of the gaseous domain. Accordingly, in the thermodynamic limit, the isotherm shows a cusp at the percolation threshold followed by a steep but continuous decrease
(the cusp is rounded in finite samples).

The simulation of the desorption process thus requires the use of an explicit external reservoir adjacent to the gel sample, which of course introduces a severe anisotropy in the model and makes it difficult to calculate radially averaged correlation functions. To circumvent this problem, we have used another procedure where the desorption is not initiated by the interface with an external  reservoir but triggered by the presence of 
gas bubbles inside the material.  We have indeed observed in our previous studies (see Fig. 16 in Ref.\cite{DKRT2003}) that the last desorption scanning curves (obtained by stopping the adsorption just before saturation and then decreasing the chemical potential) look very much like
the desorption isotherms obtained in presence of a reservoir (when averaging the
fluid density deep inside the aerogel, which gives a good estimate of the isotherm 
in the thermodynamic limit).  Near the end of the adsorption process, the remaining gaseous domain is composed of isolated bubbles which obviously play the same role as an external reservoir when the chemical potential is decreased. The advantage of initiating the  desorption with these last
bubbles is that one can use the same geometry as during adsorption, with 
periodic boundary conditions in all directions (moreover, using  
small bubbles instead of a planar interface of size $L^2$ considerably suppresses
finite-size effects).

In practice, the calculation has been performed by keeping five bubbles in each
sample. This implies that the chemical potential at which desorption is
initiated is slightly different in each sample (if one chooses the same $\mu$ in
all samples, some of them may be already completely filled with liquid). This number of bubbles results from a compromise: on the one hand, keeping a single bubble may not be sufficient  to trigger the desorption process (in some samples, the growth of the bubble is hindered by the neighboring gel particles and the desorption occurs at a much lower value of the chemical potential); on the other hand, 
keeping too many bubbles results in a too rapid growth of the gas domain.  
As can be seen in Fig. 2, the isotherm obtained with this procedure is indeed very
 close to the isotherm calculated in presence of an
external  reservoir. 

The fluid-fluid and solid-fluid correlation functions computed at several points
along the desorption branch are shown in Figs. 7 and 8.
\begin{figure}[hbt]
\includegraphics*[width=10cm]{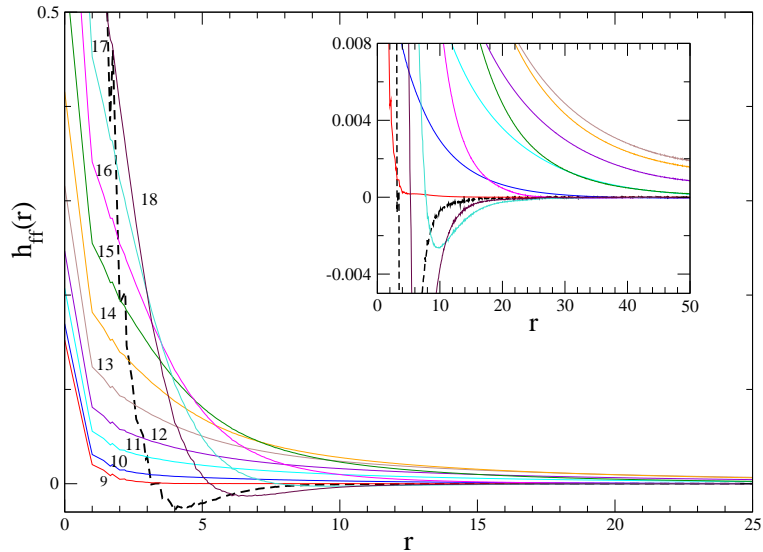}
\caption{ Fluid-fluid correlation function $h_{ff}(r)$ along the desorption isotherm
in a $87\%$ porosity aerogel at $T^*=0.5$. The numbers refer to the points
in Fig. 2; the dashed line is the gel correlation function $h_{gg}(r)$. The vertical scale is expanded in the inset,
showing the very slow decrease of $h_{ff}(r)$ towards zero in the steepest portion of the isotherm (Color on line).}
\end{figure}
\begin{figure}[hbt]
\includegraphics*[width=10cm]{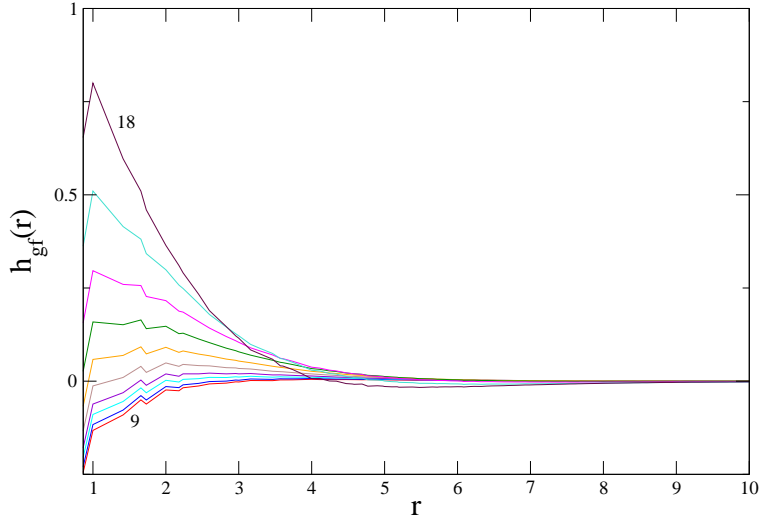}
\caption{Same as Fig. 7 for the cross correlation function $h_{gf}(r)$ (Color on line). }
\end{figure}

One can see that $h_{ff}(r)$ is dramatically changed with respect to
adsorption: from saturation down to $\mu\approx-4.7$ (curves $9$ to $16$), the function is 
monotonically decreasing, showing no minimum.  Moreover, as shown in the inset of Fig. 7, a long-range
tail is growing and  $h_{ff}(r)$ may differ significantly
from zero at $r=L/2$\cite{note10}.
Although it is difficult to
characterize this decrease by a unique and well-defined correlation length  (the curve cannot
be fitted by a simple function such as an exponential), it is clear that the range of the correlations
 is small at the beginning of
the desorption process, then increases considerably, goes through a maximum in the
steeppest portion of the isotherm (corresponding approximately to point $13$
in Fig.2), and eventually decreases. As the hysteresis loop closes and the adsorbed phase
consists again of a liquid film coating the areogel strands, a shallow minimum
reappears in the curve, which is reminiscent of the underlying gel structure.

In contrast, the changes in the cross-correlation function $h_{gf}(r)$ between
adsorption and desorption are very small (the function has only a
slightly longer tail during
desorption). It appears that the gel-fluid correlations depend 
essentially  on the average fluid density: for a given value of $\rho_f$,
they are almost the same on the two branches of the hysteresis loop.

The calculation of the fluid-fluid structure factor $S_{ff}(q)$
is complicated by the fact that $h_{ff}(r)$ decreases very slowly to zero, and one cannot
use anymore the continuation procedure that forces the  sum-rule, Eq. 11, to be satisfied
(the results for $q<2\pi/L$ change considerably with $L$, which shows that the resulting curve  is not a good approximation of the
infinite-size limit).
Instead, we have only subtracted from $h_{ff}(r)$ its value at $r=L/2$ (setting $h_{ff}=h_{ff}(r=L/2)$
in Eq. 12) so as to avoid the large but spurious oscillations in $S_{ff}(q)$ that result 
from the discontinuity at $L/2$ (there is still a discontinuity in the slope of 
$h_{ff}(r)$ at $L/2$
that  induces small oscillations in some of the curves of Fig. 9). It is clear that finite-size effects are
important  in this case and the results for $q<2\pi/L$ must be considered with caution.
\begin{figure}[hbt]
\includegraphics*[width=10cm]{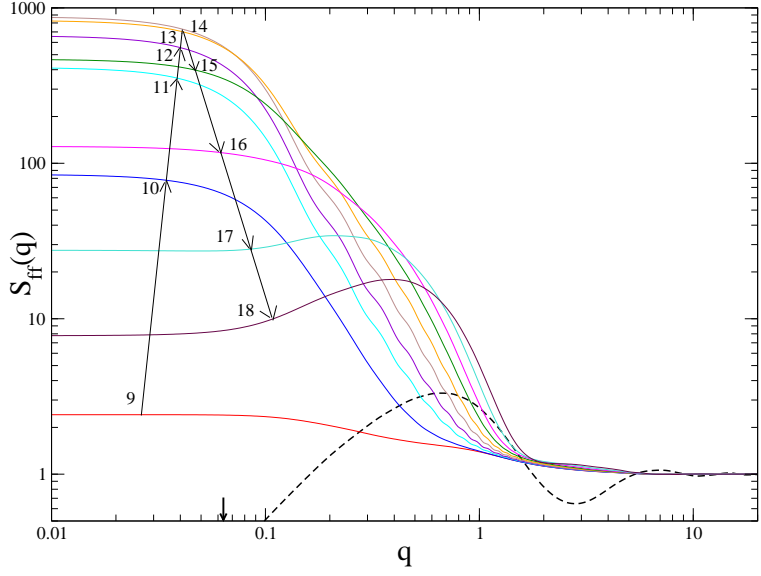}
\caption{ Fluid-fluid structure factor  $S_{ff}(q)$ along the desorption isotherm
in a $87\%$ porosity aerogel at $T^*=0.5$. The numbers and arrows refer to points $9$ to $18$
in Fig. 2. The dashed line is the gel structure factor $S_{gg}(q)$ (Color on line).}
\end{figure}

The resulting fluid-fluid structure factors  along the desorption isotherm are shown in Fig. 9. (We do not present the curves
for $S_{gf}(q)$ as they look very much like those in Fig. 6 with only a slightly broader peak.)
As could be expected, the structure factors  computed just before and after the knee
in the isotherm (for $\mu>-4.67$)
are very different from those obtained during adsorption.
Firstly, the peak  that was associated to the minimum in $h_{ff}(r)$ has now
disappeared and the small-$q$ intensity  saturates to a value that is considerably
larger than the maximum value obtained during 
adsorption (compare the vertical scales in Figs. 5 and 9). Secondly, as $\mu$ varies 
from $-4.65$ to $-4.67$ (curves $11$ to $14$), there is a linear portion
in $S_{ff}(q)$ whose maximal extension is about one decade on a
log-log scale. On the other hand,
when $\mu$ is decreased further, the peak in $S_{ff}(q)$ is recovered and the curves become more similar to
the ones  computed on the adsorption branch.

The linear regime in $S_{ff}(q)$  strongly suggests the presence of fractal
correlations\cite{note3}. However,
according to our previous studies\cite{DKRT2004}, it is only the {\it gaseous} domain that
should be a fractal object at the percolation threshold, as illustrated by the isotropic and strongly 
ramified structure shown in Fig. 10.
\begin{figure}[hbt]
\includegraphics*[width=7cm]{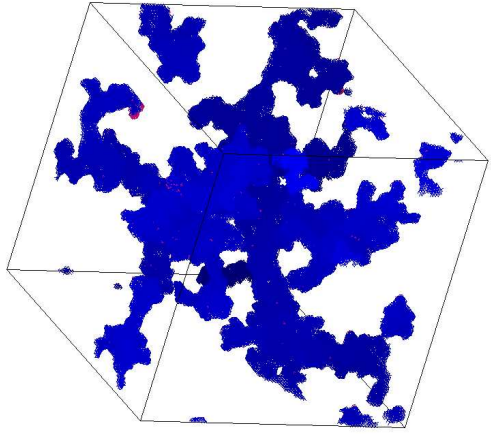}
\caption{Snapshot of the vapor domain in a $87\%$ gel sample during 
desorption at $T^*=0.5$ and $\mu=-4.63$ (Color on line).}
\end{figure}
The correlation function $h_{ff}(r)$, on the other
hand, does not discriminate between a site representing a gaseous region 
($\rho_i\approx 0$ at low
temperature) and a site representing a gel particle ($\rho_i\equiv 0$). In order to
really show the existence of fractal correlations within the gas domain, one must consider either the `gas-gas'
correlation function (defining the quantity $\rho^{gas}_i=\eta_i-\rho_i$ which is
equal to $1$ only in the gas phase) or the complementary function that measures the correlations within
the dense (solid or liquid) phase (one then defines $\rho^{dense}_i=1-\rho^{gas}_i$).
The corresponding structure factor $S_{dd}(q)$ is the quantity that is measured experimentally
when using the `contrast matching' technique\cite{H2004}. It is related to 
 $S_{ff}(q)$ and $S_{gf}(q)$ by
\begin{equation}
(\rho_g+\rho_f)(S_{dd}(q)-1)=\rho_f(S_{ff}(q)-1)+\sqrt{\rho_g\rho_f}S_{gf}(q)+\rho_g(S_{gg}(q)-1) \ .
\end{equation}
$S_{dd}(q)$ is shown in Fig. 11 for $\mu=-4.65$\cite{note11}, in the region of the
knee in the desorption isotherm (point $11$ in Fig. 2). It clearly contains a linear
 portion over almost one decade and can be very well represented in this range of wavevectors
 by the fit\cite{FKS1986}
\begin{equation}
S_{dd}(q)\sim \frac{\sin\left[ (d_f-1) \tan^{-1}(ql)\right]}
{ q \left[ l^{-2}+q^2 \right]^{(d_f-1)/2}}
\end{equation}
with $d_f=2.45$ and $l=17$, where $l$ is a crossover length that limits the fractal
regime at large distances. Note that the linear portion  itself 
has a slope $\approx -2.1$ (the above formula reproduces the right slope only
when $l$ is very large\cite{H1994}). An accurate determination of  $d_f$  would
therefore require a much larger system and, at this stage, it is not possible to 
decide if the fractal dimension is consistent with that of random percolation. In any 
case, these results strongly
suggest that the gas domain exhibits fractal correlations during desorption, correlations which have
no relation with the underlying gel microstructure (we recall that there is almost no fractal regime in 
the $87\%$ aerogel, as can be seen in Fig. 1).
\begin{figure}[hbt]
\includegraphics*[width=8cm]{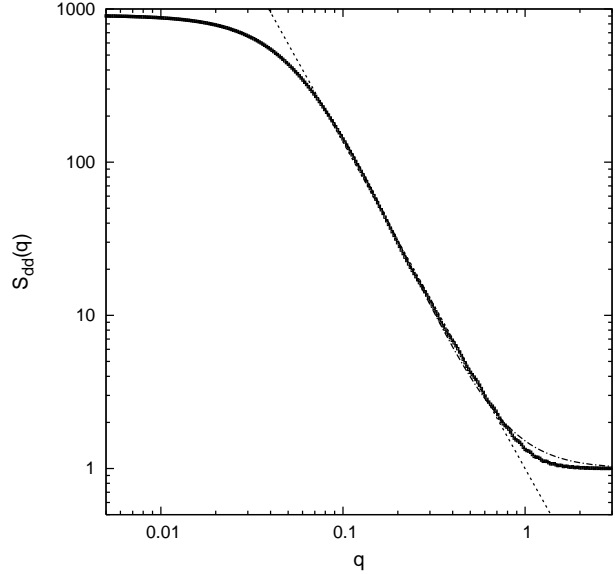}
\caption{Structure factor of the dense phase (see text), $S_{dd}(q)$, during desorption at $T^*=0.5$ and
$\mu=-4.65$. The dashed-dotted curve is the best fit according to Eq. 16 and the straight dashed
line has a slope $-2.1$.}
\end{figure}

Raising the temperature to $T^*=0.8$ has a dramatic effect, as shown in Fig.12.
The maximum value of $S_{ff}(q)$ has dropped by two orders of magnitude  and
there is no significant region with a fractal-like power-law behavior. Indeed, $h_{ff}(r)$ has no
more a long-range tail and the correlations are very
similar during adsorption and desorption, as could be expected from the very
thin shape of hysteresis loop.
This is the signature that the desorption mechanism has changed, in agreement with the 
analysis of Refs.\cite{DKRT2003,DKRT2004}.
It is now due to a  cavitation phenomenon in which gas bubbles first appear
 in the largest cavities of the gel and then grow and coalesce until the whole 
 void space is invaded\cite{note13}. 
\begin{figure}[hbt]
\includegraphics*[width=9cm]{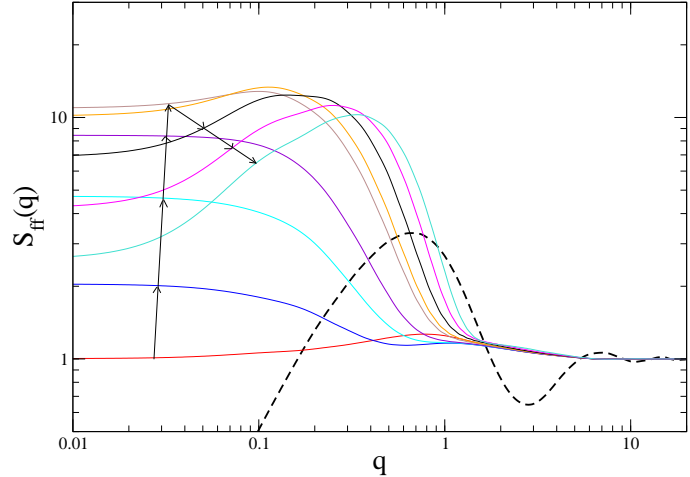}
\caption{ Fluid-fluid structure factor  $S_{ff}(q)$ along the desorption isotherm
in a $87\%$ porosity aerogel at $T^*=0.8$. (Color on line).}
\end{figure}

We have not studied the correlations during desorption in the $95\%$ porosity gel.
At very high temperature ($T^*\gtrsim 0.9$), desorption is expected to be due again
to cavitation\cite{DKRT2004}, and the results shoud be similar to those in the $87\%$
that have just been described. On the other hand, at low temperature (e.g. $T^*=0.5$), 
the theory predicts a depinning transition in which a self-affine interface sweeps through the whole
sample, resulting in a discontinuous desorption isotherm\cite{DKRT2004}. Therefore, like
in the case of the macroscopic avalanche during adsorption,  the correlations along the isotherm cannot
be studied within the framework of a grand-canonical calculation. At intermediate temperatures,
one could probably observe again  extended fractal correlations associated to a percolating cluster of gas, but
this study requires the use of larger systems so as to probe smaller values of $q$ and discriminate
the  effects due to the own fractal structure of the gel.
 
\section{Scattered intensity and comparison with experiments}

As mentionned in the introduction, there  have been  two recent scattering studies of  gas condensation in aerogels, both with 
$^4$He\cite{LMPMMC2000,LGPW2004}.
In Ref.\cite{LGPW2004}, light scattering is used to study adsorption and desorption in a $95\%$ porosity gel
at several temperatures between $4.47K$ ($T^*\approx0.86$) and $5.08K$ ($T^*\approx0.98$).  These  experimental results 
cannot be directly  compared to our theoretical predictions, and our system size is too small to investigate  the large-scale
inhomogeneities that are seen in the experiments (some of them are visible to the eye). However, there are two key observations that 
appear to be in  agreement with our predictions:
i) at the lowest temperature studied,  the optical signal due to helium adsorption is larger than if the fluid density was simply correlated to the density of silica, indicating that the correlations within the fluid extend beyond the aerogel correlation length, and  ii) the aerogel is much brighter during desorption, indicating that the characteristic size of the density fluctuations is much larger than during adsorption. 

This latter conclusion was also reached from  the small-angle x-ray scattering measurements (SAXS) performed  in a
$98\%$ porosity aerogel at $3.5K$ ($T^*\approx0.67$)\cite{LMPMMC2000}.  SAXS is particularly well suited for observing the structural features associated to fluid adsorption, and in order to compare more easily to experiments we shall push further our calculations and compute the resulting scattered intensity.  Of course, the predictions must be taken with a grain of salt, considering the limitations in the model and the theory.  
Since the scattered intensity is proportional to the Fourier transform of the electron density fluctuations, one has
\begin{align}
I(q)\propto  \rho_g F(q)^2S_{gg}(q) + 2\sqrt{\rho_g\rho_f}\alpha F(q) S_{gf}(q) + \rho_f \alpha^2S_{ff}(q)
\end{align}
where $F(q)$ is the form factor of silica particles (see section IIIA) and $\alpha$ is the ratio of the electron density in the adsorbed fluid to that in the solid. As an additional  approximation, we shall  take $F(q)\approx 1$, restricting the study to the range $2\pi/L\leq q\leq 2$ where this is presumably a reasonable approximation (in real units this corresponds to $0.02\lesssim q \lesssim 0.7$ nm$^{-1}$, taking $a=3$nm). Assuming that the adsorbed liquid has the same density as the bulk liquid  at $P_{sat}$,  and using the tabulated densities of helium and silica, one finds that $\alpha$ varies from $6.55 \ 10^{-2}$ at $T^*=0.5$ to 
$5.75 \ 10^{-2}$ at $T^*=0.8$.  

The theoretical  scattered intensities during adsorption and desorption in the  $87\%$ gel at $T^*=0.5$ are shown in Figs. 13 and 14. Like in the experiments\cite{LMPMMC2000}, we plot the ratio $R(q)=I(q)/I_e(q)$ 
where $I_e(q)$ is the contribution of the empty aerogel (the first term in the  right-hand side of Eq. 17) in order to accentuate the effects of the adsorbed  gas, especially in the initial stage of adsorption. Moreover, we hope that this also partially corrects the small-$q$  defects due to the absence of large-scale fluctuations in the DLCA gel structure factor.
\begin{figure}[hbt]
\includegraphics*[width=10cm]{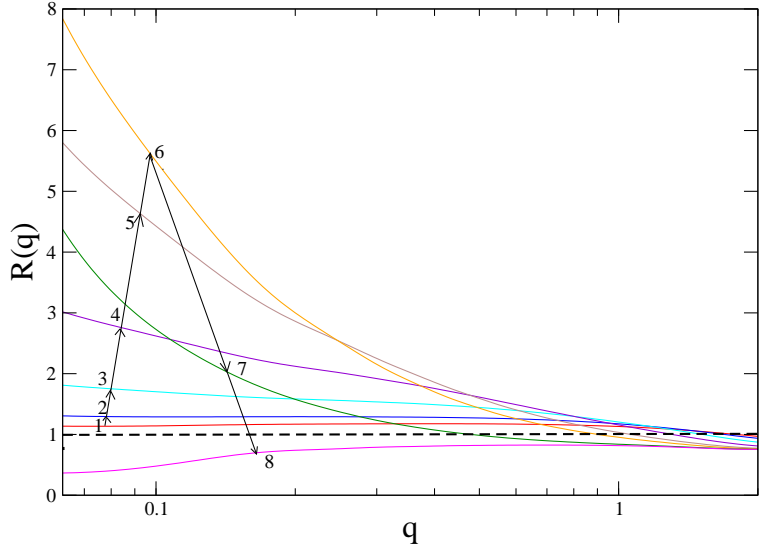}
\caption{ Theoretical ratio $R(q)$ of the scattered intensity $I(q)$ to the scattered intensity $I_e(q)$ of the empty aerogel during helium adsorption in a $87\%$  aerogel
at $T^*=0.5$. The dashed line coresponding to $R=1$ is shown for reference (Color on line).}
\end{figure}
\begin{figure}[hbt]
\includegraphics*[width=10cm]{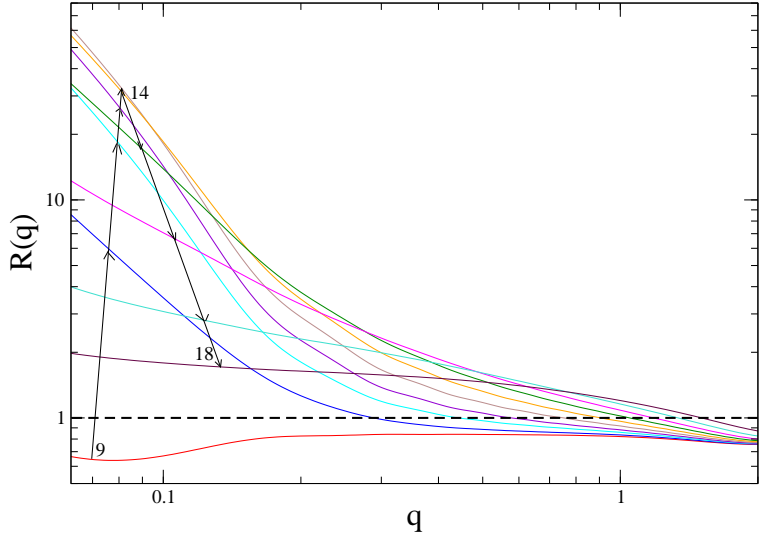}
\caption{ Same as Fig. 13 during desorption (Color on line).}
\end{figure}

The main features of the curves displayed in Fig.  13 are the following: i) At the very beginning of adsorption, $R(q)$ slightly increases but 
remains almost independent of $q$. This is the signature of the $^4$He film coating the aerogel. In this regime, the main contribution to the scattered intensity comes from the gel-fluid correlations (the second term in the right-hand side of Eq. 17). ii)  As $\mu$ increases,  the scattering grows in intensity at small  $q$, reflecting the presence of  the broad peak in $S_{ff}(q)$  that moves towards smaller wavevector with filling (see Fig. 5). iii) As the aerogel fills further, $R(q)$ decreases until it becomes again almost flat at complete filling. The total intensity is then reduced with respect to that of the empty aerogel.
Direct comparison with the experimental results of Ref.\cite{LMPMMC2000} is again problematic: the adsorption isotherm in the 
$98\%$ gel is indeed very steep at $3.5K$, suggesting that one may be in the regime of a macroscopic avalanche. However, the behavior of the experimental $R(q)$ is remarkably similar to what have just been described (preliminary measurements in a $86\%$ aerogel\cite{M2005}
also show the same trends).  The results of Ref.\cite{LMPMMC2000} were interpreted according to a model of two-phase coexistence, with a 
`film' phase in equilibrium with a filled `pore'  phase. This is at odds with the  theoretical scenario discussed in  Refs.\cite{DKRT2003,DKRT2004,DKRT2005}
which emphasizes the nonequilibrium character of the transition.  The present results seem to show that this  approach can also 
elucidate (at least qualitatively) the behavior of the scattered intensity. 

During desorption, the most characteristic feature of the curves shown in Fig. 14 is the very significant increase of the ratio $R(q)$ at small $q$ with respect  to adsorption (note the logarithmic scale on the vertical axis).  This is related to the corresponding increase in $S_{ff}(q)$ shown in Fig. 9 and is clearly due to the presence of long-range correlations within the fluid. As the desorption proceeds, $R(q)$ goes through a maximum and then decreases  until it becomes flat again. Remarkably, no power-law fractal regime is visible in $R(q)$ in the range $0.06\lesssim q\lesssim 1$ as was the case with $S_{dd}(q)$ in Fig. 11. It is the small value of $\alpha$ (due to the small electron density of He), and not the division by $I_e(q)$, which  is responsible for this unfortunate disappearance ($I(q)$ becomes proportional to $S_{dd}(q)$ when $\alpha=1$, which is only the case in a contrast matching experiment).  In the measurements of Ref.\cite{LMPMMC2000}, this increase of $R(q)$ at small $q$ is not mentionned, but the analysis of the data shows that the charateristic size of the inhomogeneities is much larger than during adsorption, as already mentionned, and that it decreases rapidly in the last stage of desorption. 
\begin{figure}[hbt]
\includegraphics*[width=10cm]{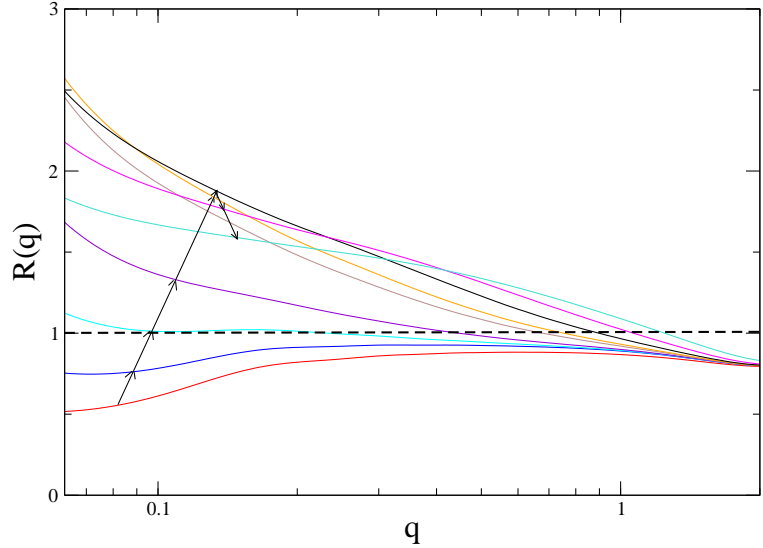}
\caption{ Same as Fig. 13 during desorption at $T^*=0.8$ (Color on line).}
\end{figure}

Not surprisingly, the theoretical scattered intensity in the $87\%$ aerogel is considerably smaller at high temperature, as illustrated in Fig. 15
for $T^*=0.8$.  The intensity ratio $R(q)$ has been divided by about $40$.  
We therefore conclude that the magnitude of the scattered intensity can indicate that 
the nature of the desorption process has changed.
 We leave to our colleague experimentalists the challenge of checking the presence of a fractal-regime during desorption, as was done in Vycor\cite{LRHHI1994,PL1995,KKSMSK2000} and xerogel\cite{H2002}.

\acknowledgments
We are grateful to N. Mulders for very useful discussions and communication of unpublished results.
The Laboratoire de Physique Th\'eorique de la Mati\`ere Condens\'ee is the UMR 7600 of
the CNRS.


\begin{thebibliography}{10}
\bibitem{note5} On the other hand, a box of size $L=200$ has been used to simulate
the aerogels with porosity  $\phi\geq0.97$.
\bibitem{note6} According to Babinet principle, two `complementary' objects have
 the same diffraction properties. When the gel is completely filled with a liquid
 whose density is approximately uniform and close to $1$ (which is the case
 at low temperature), Babinet
 principle (in real space) implies that
 $h_{ff}(r)\approx (1-\phi)^2/\phi^2 \ h_{gg}(r)$ and  $h_{gf}(r)\approx
 -(1-\phi)/\phi \ h_{gg}(r)$. This is not quite visible in Figs. 3(b) and 4(b) because
 of the drastic change in the vertical scales.
\bibitem{note7} We recall that $S_{ff}(q)$ has been computed so as to satisfy Eq. 11,
the correct $q=0$ limit of $h_{ff,d}(q)$. Then  $S_{ff}(q)$
 is expected to represent more faithfully the structure factor of the infinite 
 system. In fact, setting $h_{ff}=0$ or $\neq 0$ in Eq. 12 only changes $S_{ff}(q)$
for $q\lesssim2\pi/L$ and the differences are small, as illustrated by the
dotted curve in Fig. 5. The same is true for $S_{gf}(q)$.
\bibitem{note8} One could imagine to study separately the samples in which the jump has not yet occured and those which are already in
the high-density phase, and then to use a sort of  `lever'  rule to compute the structural properties at intermediate densities. However, this would be an incorrect procedure: the two fluid configurations before and after the avalanche are not in equilibrium (they are both metastable and the grand-potential has decreased during the avalanche) and one cannot apply to this problem the standard methods of phase equilibria.
\bibitem{note10} This can be understood by noting that Eq. 10 can be
rewritten as $\rho_f^2 h_{ff,d}({\bf r})=(1/N)\sum_{i,j}
[\overline{\delta\rho_i\delta\rho_j}
+\overline{\rho_f^2(\{\eta_i\})}-\rho_f^2]\
\delta_{{\bf r},{\bf r}_{ij}}$ where $\delta\rho_i=\rho_i-\rho_f(\{\eta_i\})$
represents the deviation of $\rho_i$ from the average density in the sample. The term $\overline{\rho_f^2(\{\eta_i\})}-\rho_f^2$ is
of order $1/N$ but it may be not negligible in a finite system when
the number of fluid particles fluctuates significantly from one sample to another.
It may then dominate the long-range behavior of $h_{ff,d}({\bf r})$.
This is precisely what happens in the steepest
 portion of the desorption isotherm where
 some samples are still filled with liquid whereas other ones are already significantly
 drained.
\end{thebibliography}
\end{document}